# Bootstrap p-values reduce type 1 error of the robust rank-order test of difference in medians

*Nirvik Sinha\*,a,b*

*aNorthwestern Interdepartmental Neuroscience Program, Feinberg School of Medicine, Northwestern University, 320 E Superior Street, Morton 1-645, Chicago, IL 60611-3010, USA*

*bDepartment of Physical Therapy and Human Movement Sciences, Feinberg School of Medicine, Northwestern University, 645 N. Michigan Ave., Suite 1100, Chicago, IL 60611, USA*

*\* corresponding author: Nirvik Sinha: nirviksinha@gmail.com*

**Abstract**

The robust rank-order test (Fligner and Policello, 1981) was designed as an improvement of the non-parametric Wilcoxon-Mann-Whitney U-test to be more appropriate when the samples being compared have unequal variance. However, it tends to be excessively liberal when the samples are asymmetric. This is likely because the test statistic is assumed to have a standard normal distribution for sample sizes > 12. This work proposes an on-the-fly method to obtain the distribution of the test statistic from which the critical/p-value may be computed directly. The method of likelihood maximization is used to estimate the parameters of the parent distributions of the samples being compared. Using these estimated populations, the null distribution of the test statistic is obtained by the Monte-Carlo method. Simulations are performed to compare the proposed method with that of standard normal approximation of the test statistic. For small sample sizes ($<= 20$), the Monte-Carlo method outperforms the normal approximation method. This is especially true for low values of significance levels ($< 5\%$). Additionally, when the smaller sample has the larger standard deviation, the Monte-Carlo method outperforms the normal approximation method even for large sample sizes (= 40/60). The two methods do not differ in power. Finally, a Monte-Carlo sample size of $10^4$ is found to be sufficient to obtain the aforementioned relative improvements in performance. Thus, the results of this study pave the way for development of a toolbox to perform the robust rank-order test in a distribution-free manner.

# Introduction

Experiments designed to compare the central tendencies of the variable of interest between two independent conditions often generate response measurements that can be ordered (i.e. ranked), but the location of the response on a scale of measurement is arbitrary. The ubiquity of this phenomenon entails the extensive use of non-parametric statistical methods in experimental research (e.g. in the field of social sciences and consumer research). Moreover, practical issues such as failure to satisfy the assumptions of normality and homoscedasticity or small sample sizes further invalidate the use of parametric statistics in such cases.

The Wilcoxon-Mann-Whitney U-test is the most employed non-parametric test for comparing independent samples [1, 2]. However, when the sample standard deviations are unequal, this test does not always maximize the correct decisions. For example, when the smaller sample has greater standard deviation, the type 1 error (i.e. the false positive rate) increases irrespective of whether the t-test or the Wilcoxon-Mann-Whitney U-test is used. In fact, when the sample sizes are equal (with unequal standard deviation) or when the smaller sample has greater standard deviation, the t-test is more powerful as compared to the Wilcoxon-Mann-Whitney U-test [3]. One way of dealing with this problem is to employ the Welch's t-test (which corrects for unequal sample standard deviations) in conjunction with the rank transformation of Wilcoxon-Mann-Whitney U-test [4]. Sen obtained a distribution-free estimate of the variance of the Wilcoxon-Mann-Whitney U-test statistic and utilized this to attach confidence bounds to $P(X < Y)$ where $X$ and $Y$ are the test samples [5]. In this work the approach due to Fligner and Policello is adopted. These researchers modified the Wilcoxon-Mann-Whitney U-test to account for unequal standard deviations and devised what they termed as the 'robust rank-order test' [6]. However, the robust rank-order (RRO) test, despite its proven advantage over the Wilcoxon-Mann-Whitney U-test, has not achieved widespread use for comparison of non-normal samples with unequal standard deviations. As pointed out in [7], one important reason for this may be due to the fact that there is limited knowledge about the distribution of the RRO test statistic, especially its critical values for various levels of significance. Fligner and Policello computed the exact critical values for levels of significance $\alpha = 0.01, 0.05, 0.025$, and $0.01$, and for sample sizes up to 12 [6, 8]. For larger sample sizes, they recommended that the test statistic be approximated by a standard normal distribution. However, the convergence of the test statistic to the standard normal variable with increase in sample size is rather slow. Even for samples of the size 40 normal approximation of the test statistic results in elevated Type 1 error [9]. To address this issue, Feltovich expanded on the list of available critical values (for up to $m, n \leq 40$, where $m$ and $n$ are the sample sizes). Exact critical values were provided for $m$ up to 6 and $n$ up to 40, for $m = 7$ and $n$ up to 30, for $m = 8$ and $n$ up to 23, for $m = 9$ and $n$ up to 20, for $m = 10$ and $n$ up to 18, for $m = 11$ and $n$ up to 16, for $m = 12$ and $n$ up to 14, and for $m = n = 13$. For the remaining values of $m$ and $n$, approximate critical values were reported based on 1 million Monte-Carlo pairs of samples drawn from the standard uniform distribution [7]. Use of these computed critical values for comparing equal or unequal sized samples with unequal standard deviations reduced the type 1 error. On the other hand, comparison of samples drawn from different asymmetric distributions with unequal standard deviations using the computed critical values yielded a type 1 error which was only marginally lower than that obtained by normal approximation of the test statistic [9].

Given the poor performance of the RRO test on samples drawn from skewed distributions, the objective of this work was to improve the technique of critical value estimation in such cases using the Monte-Carlo method so that there is a significant reduction of the type 1 error as compared to when the test statistic is approximated using the standard normal distribution. Furthermore, this technique should be generalizable to the cases when the samples are drawn from various kinds of skewed distributions. Accordingly, the method of maximum likelihood estimation was used to approximate the parent distributions from which the samples were drawn. Then the critical values were estimated using these approximated distributions (after equalizing their central tendencies) with the Monte-Carlo method. The RRO test performed using these critical values was shown to exhibit lower type 1 error as compared to when the test statistic was assumed to be normally distributed. This was also achieved with minimal reduction in power of the test. Finally, the minimum number of Monte-Carlo sample pairs required to obtain these results was estimated so that the critical value of the test statistic (for any particular significance level) may be computed on the fly without the aid of any look-up tables (for up to $m, n > 13$).

## Methods

### The robust rank-order test

Let $X$ and $Y$ be two independent random samples (of sizes $m$ and $n$) drawn from their parent populations with probability density functions $F(x, \varphi_x)$ and $G(y, \varphi_y)$ respectively (where $\varphi_x$ and $\varphi_y$ are the population parameters). If $\theta_x$ and $\theta_y$ are the sample medians (assumed to be unique) of $X$ and $Y$, then the problem is to test the null hypothesis $H_0$: $\theta_x = \theta_y$ against the alternative hypothesis $H_1$: $\theta_x \neq \theta_y$ (or one of the one-sided alternatives). Note that here we are testing for the equality of the location parameters without assuming equal standard deviations or similar shapes of the two parent distributions. Given this problem, we perform the RRO test as follows [6]: The *placement* of each element $x_i$ in $X$ and $y_i$ in $Y$ is computed as the number of lower-valued observations in the other sample (i.e. $Y$ and $X$). Let $U(Y, X_i)$ and $U(X, Y_i)$ be the respective placements. The *mean placements* of $X$ and $Y$ is obtained as the arithmetic means of $U(Y, X_i)$ and $U(X, Y_i)$. Let them be denoted by $U(Y, X)$ for $X$ and $U(X, Y)$ for $Y$. Then an index of variability of the *placements* is computed as follows:

$$V_x = \sum_{i=1}^{m}[U(Y, X_i) - U(Y, X)]^2 \tag{1a}$$

$$V_y = \sum_{i=1}^{n}[U(X, Y_i) - U(X, Y)]^2 \tag{1b}$$

The RRO test statistic $Ú$ is given by:

$$Ú = \frac{m.U(Y,X) - n.U(X,Y)}{2\sqrt{V_x + V_y + U(X,Y).U(Y,X)}} \tag{2}$$

### Estimation of the critical value for the test statistic $Ú$

To estimate the critical value for $Ú$ the following procedure is adopted:

a. Generate $Y'$ from $Y$ such that it has a central tendency equal to $X$ [10]:

$$Y' = Y - \Delta \tag{3}$$

where $\Delta$ is the Hodges-Lehman shift operator:

$$\Delta = median\ [Y_j - X_i\ (1 \leq i \leq m, 1 \leq j \leq n)] \tag{4}$$

b. Estimate the parameters $\varphi_x$ and $\varphi_{y'}$ of the parent populations of $X$ and $Y'$ by maximizing their log-likelihood functions:

$$\widehat{\varphi_x} = \arg max\ [L(\varphi_{x,i}, X)]_{\varphi_{x,i}\ \epsilon\ \Phi_x} \tag{5a}$$

$$\widehat{\varphi_{y'}} = \arg max\ [L(\varphi_{y',i}, Y)]_{\varphi_{y',i}\ \epsilon\ \Phi_{y'}} \tag{5b}$$

where $\widehat{\varphi_x}$ and $\widehat{\varphi_{y'}}$ are the estimates of $\varphi_x$ and $\varphi_{y'}$, respectively. The parameter spaces $\Phi_x$ and $\Phi_{y'}$ are the set of real valued $d_x$ and $d_{y'}$-dimensional vectors ($\Phi_x \epsilon\ \mathbb{R}^{d_x}$, $\Phi_y \epsilon\ \mathbb{R}^{d_y}$) whose elements define the parent distributions $F$ and $G$. Given the parameters $\varphi_{x,i}$ and $\varphi_{y',i}$, the log-likelihood is calculated as the logarithm of the product of the corresponding marginal probabilities of $X$ and $Y$:

$$L(\varphi_{x,i}, X) = \log_e \prod_{i=1}^{m} F(x_i, \varphi_{x,i}) \tag{6a}$$

$$L(\varphi_{y',i}, Y') = \log_e \prod_{i=1}^{n} G(y'_i, \varphi_{y',i}) \tag{6b}$$

c. Generate the null distribution $\Xi$ of $\acute{U}$ by computing $\acute{U}$ for randomly drawn sample pairs from the estimates of the parent populations (i.e. with parameters $\widehat{\varphi_x}$ and $\widehat{\varphi_{y'}}$).

d. For a given p-value $p$ (left/right/two-tailed), compute the critical value as:

a. Right-tailed: $\acute{U}_{cR} = \arg max[\frac{n(\Xi \geq \acute{U}_\iota)}{n(\Xi)}]_{i=1}^{n(\Xi)} \mid [\frac{n(\Xi \geq \acute{U}_\iota)}{n(\Xi)} \leq p]$ (7a)

b. Left-tailed: $\acute{U}_{cL} = \arg max[\frac{n(\Xi \leq \acute{U}_\iota)}{n(\Xi)}]_{i=1}^{n(\Xi)} \mid [\frac{n(\Xi \leq \acute{U}_\iota)}{n(\Xi)} \leq p]$ (7b)

c. Two-tailed: $\acute{U}_{cL} = \arg max[\frac{n(\Xi \leq \acute{U}_\iota)}{n(\Xi)}]_{i=1}^{n(\Xi)} \mid [\frac{n(\Xi \leq \acute{U}_\iota)}{n(\Xi)} \leq \frac{p}{2}]$ (7c)

$\acute{U}_{cR} = \arg max[\frac{n(\Xi \geq \acute{U}_\iota)}{n(\Xi)}]_{i=1}^{n(\Xi)} \mid [\frac{n(\Xi \geq \acute{U}_\iota)}{n(\Xi)} \leq \frac{p}{2}]$ (7d)

where $\acute{U}_{cR}$ and $\acute{U}_{cL}$ are the estimated critical values of the test statistic for the right and left-tailed p-values, respectively and $n(A)$ is the cardinality of the set $A$. Alternatively, the actual p-value for the original test statistic ( $\acute{U}_o$, computed from $X$ and $Y$) can be calculated as:

a. Right-tailed p-value, $p_L = \frac{n(\Xi \geq \acute{U}_o)}{n(\Xi)}$ (8a)

b. Left-tailed p-value, $p_R = \frac{n(\Xi \leq \acute{U}_o)}{n(\Xi)}$ (8b)

c. Two-tailed p-value, $p = \min\ (2.\min(p_L, p_R), 1)$ (8c)

The main assumption of this estimation method is that the probability density function of the parent distributions is known. This is of course not the case in real world samples. One practical solution is to iterate the likelihood maximization step $b$ over a predefined set of distributions with known probability density functions and choose the distribution which best fits the data of each sample. Another potential issue with maximum likelihood estimation is what should be the starting values of the parameters to be estimated for fitting the data to a distribution? One way to ensure consistently good starting values is to use the parameters of the distribution which best fit the 1$^{st}$ 3 or 4 moments of the distribution (either analytically if possible or by numerical optimization).

## Simulations and Results

To illustrate the improvement in performance of the proposed method of estimation of the distribution of the test-statistic (relative to its normal approximation), a set of 9 simulations were conducted in MATLAB R2020a using synthetically generated datasets. Since the focus of this study was samples with skewed distributions, the Johnson's $S_U$ distribution was used to create the set of asymmetric samples for testing. If $z$ is a standard random variable then Johnson's $S_U$ transformation is defined as [11]:

$$r = \lambda \sinh\left(\frac{z-\gamma}{\delta}\right) + \xi \tag{9}$$

where $\lambda, \gamma, \delta$ and $\xi$ are real-valued parameters such that $\delta > 0$ and $\gamma > 0$. The probability density function of the distribution is given by [11]:

$$F(x, \lambda, \gamma, \delta, \xi) = \frac{\delta}{\lambda\sqrt{2\pi}} \cdot \frac{1}{\sqrt{1+\left(\frac{x-\xi}{\lambda}\right)^2}} e^{-\frac{1}{2}\left(\gamma+\delta\sinh^{-1}\left(\frac{x-\xi}{\lambda}\right)\right)} \tag{10}$$

Each of the subsequently described simulation sets were conducted using 100,000 pairs of samples. Equal-sized sample pairs of sizes 15,20, 40 and 60 were used along with unequal-sized sample pairs of sizes $(m, n) = (15,20), (20,40), (40,60)$. In those cases where unequal-sized sample pairs had unequal standard deviations or their parent distributions were not the same, the simulations were repeated for both sizes $(m, n)$ and $(n, m)$. Two types of parent distributions were used to generate the samples i.e. left and right-skewed distributions. For this purpose, the median, standard deviation, and the sample skewness were first defined. Then the corresponding $S_U$ distribution parameters were computed using MATLAB's global optimization method *patternsearch* (with the inequality constraints $\delta > 0$ and $\gamma > 0$). In the objective function provided to the algorithm, the central moments were computed on a large sample (of size 10,00,000) drawn from the $S_U$ distribution with the current given parameters and the unweighted sum of the L2 errors of the central moments were returned for minimization. To ensure deterministic sampling in the objective function, MATLAB's random number generator was initialized to the Mersenne Twister generator with seed 0 before the sample was drawn. Additionally, to ensure repeatability and uniformity in all the Monte-Carlo simulations, before each optimization and sample-set generation, MATLAB's random number generator was initialized to the Mersenne Twister generator with seed 0. The cardinality of the null distribution of $Ú$ in the RRO test was 100,000 in all cases (see simulation set 8 to examine how the performance of the RRO test varies with the choice of this value).

In all the simulation sets where the medians of the sample pairs were equal (and hence the null hypothesis was correct), the excess type-1 error above its nominal value (as per the significance level) was computed to compare the performance of the RRO test using the estimated distribution of the test statistic vs. its standard normal approximation. Therefore, a positive/negative excess type-1 error occurs when the test is liberal/conservative respectively at the corresponding significance level. In simulation sets 3A and 3B (described below), where the medians of the sample pairs were unequal (hence the alternative hypothesis is correct), the power $(1 -$ false-negative rate) of the RRO test was computed for the same purpose. In each case the type-1 error/power was computed for 1-10% significance levels in steps of 1%. The parameters of all the simulation sets are summarized in Table 1.

Table 1: Simulation parameters

| Simulation set no. | Parent distribution name | Parent distribution central moments (median, standard deviation, skewness) | Sample sizes (S1, S2)* |
|---|---|---|---|
| 1 | i. Left-skewed<br>ii. Right-skewed | (0.0, 1.0, -1.5)<br>(0.0, 1.0, 1.5) | Equal: 15, 20, 40, 60<br>Unequal: (15,20), (20,40), (40,60) |
| 2 | i. Left-skewed<br><br>ii. Right-skewed | S1: (0.0, 1.0, -1.5)<br>S2: (0.0, 2.0, -1.5)<br><br>S1: (0.0, 1.0, 1.5)<br>S2: (0.0, 2.0, 1.5) | Equal: 15, 20, 40, 60<br>Unequal: (15,20), (20,15), (20,40), (40,20), (40,60), (60,40) |
| 3A<br>[ES = 0.25]** | i. Left-skewed<br><br>ii. Right-skewed | S1: (0.0, 1.0, -1.5)<br>S2: (0.3723, 1.0, -1.5)<br><br>S1: (0.0, 1.0, 1.5)<br>S2: (0.3793, 1.0, 1.5) | Equal: 15, 20, 40, 60<br>Unequal: (15,20), (20,40), (40,60) |
| 3B<br>[ES = 0.5]** | i. Left-skewed<br><br>ii. Right-skewed | S1: (0.0, 1.0, -1.5)<br>S2: (0.8061, 1.0, -1.5)<br><br>S1: (0.0, 1.0, 1.5)<br>S2: (0.8168, 1.0, 1.5) | Equal: 15, 20, 40, 60<br>Unequal: (15,20), (20,40), (40,60) |
| 4 | Right-skewed | S1: (0.0, 1.0, 1.0)<br>S2: (0.0, 1.0, 1.5) | Equal: 15, 20, 40, 60<br>Unequal: (15,20), (20,15), (20,40), (40,20), (40,60), (60,40) |

| 5 | Right-skewed | S1: (0.0, 1.0, 1.0) S2: (0.0, 2.0, 1.5) | Equal: 15, 20, 40, 60 Unequal: (15,20), (20,15), (20,40), (40,20), (40,60), (60,40) |
|---|---|---|---|
| 6 | Left-skewed | S1: (0.0, 1.0, -1.0) S2: (0.0, 1.0, 1.5) - | Equal: 15, 20, 40, 60 Unequal: (15,20), (20,15), (20,40), (40,20), (40,60), (60,40) |
| 7 | Left-skewed | S1: (0.0, 1.0, -1.0) S2: (0.0, 2.0, -1.5) | Equal: 15, 20, 40, 60 Unequal: (15,20), (20,15), (20,40), (40,20), (40,60), (60,40) |
| 8 | i. Left-skewed ii. Right-skewed | (0.0, 1.0, -1.5) (0.0, 1.0, 1.5) | Equal: 15 |

*S1 = sample 1, S2 = sample 2; **ES = effect size.*

**Simulation set 1**

In this simulation set the sample pairs were drawn from the same distribution type (repeated for left-skewed and right-skewed distributions separately, see Table 1) with identical median, standard deviation, and skewness. The RRO test was performed on each sample pair using the Monte-Carlo and normal approximation methods and the resultant two-tailed p-values were computed using equation 8c. For each set of 100,000 simulations (of a particular sample-size pair), the excess type-1 error was computed as:

$$excess\ type\text{-}1\ error\ (\%) = \left(\frac{n(P<\alpha)}{n(P)} - \alpha\right) * 100 \tag{11}$$

where $P$ is the set of computed p-values and $\alpha$ is the significance level. Figure 1 shows the excess type-1 errors for the RRO test using the Monte-Carlo vs. normal approximation method. The RRO test using the Monte-Carlo method shows lower excess type 1 error as compared to the normal approximation method. The difference is prominent for smaller sample sizes ($m = n = 15, 20$) and increases as the level of significance decreases. Additionally, the results are similar for equal vs. unequal-sized sample pairs as well as for left vs. right-skewed distributions.

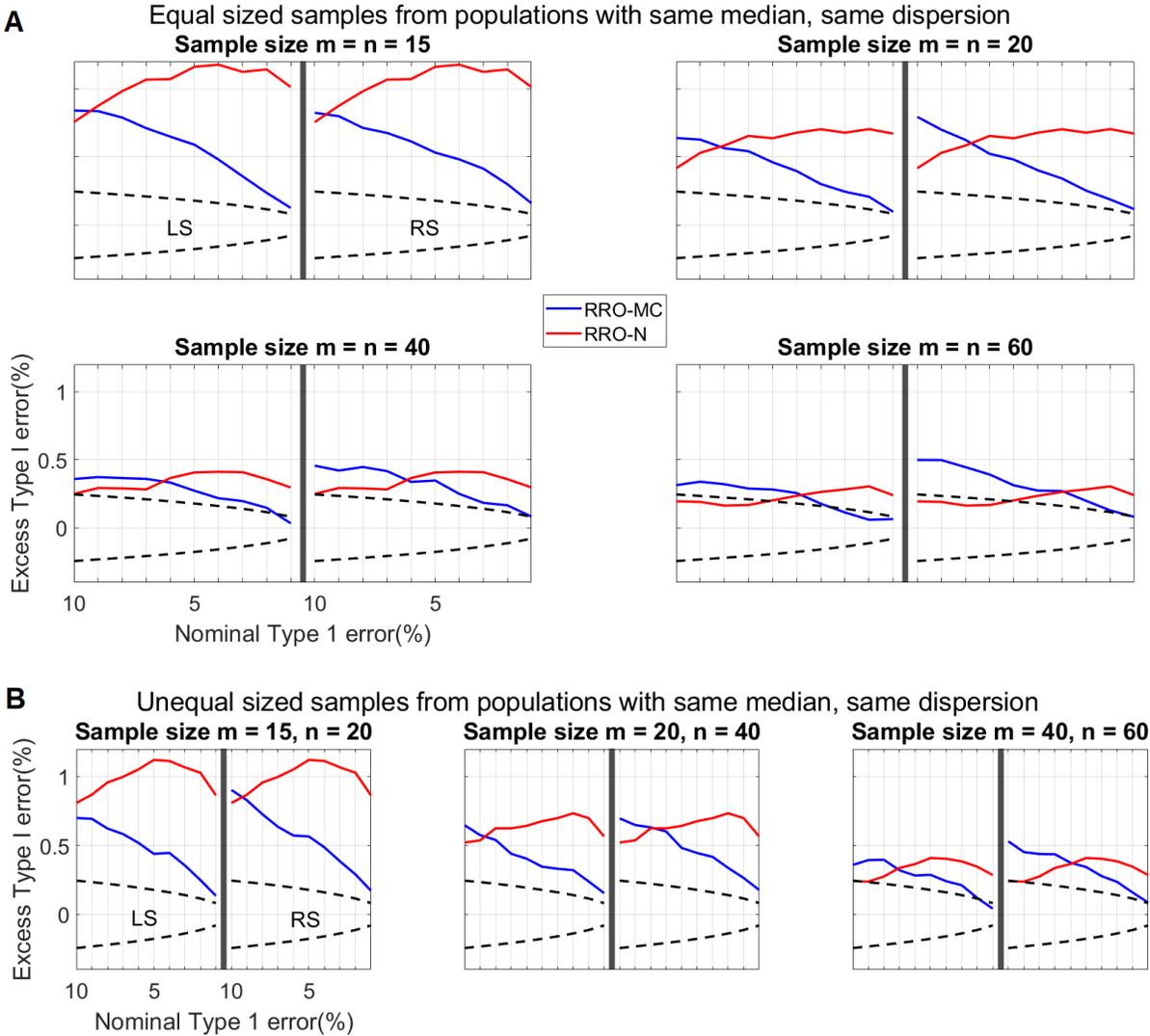

**Figure 1**: Excess type 1 error of the robust rank-order test for sample pairs drawn from the same distribution type (left/right-skewed) with identical median, standard deviation, and skewness. [A] Equal-sized sample pairs, [B] Unequal-sized sample pairs. The dashed lines enclose the region where the excess type 1 error is not significant (binomial test, p = 0.01). [LS: Left-skewed, RS: Right-skewed, RRO-MC: Robust rank-order test with Monte-Carlo estimation of test-statistic distribution, RRO-N: Robust rank-order test with standard normal approximation of test-statistic distribution]

**Simulation set 2**

This set of simulations is the same as simulation set 1 except that the standard deviations of the sample pairs are unequal (specifically, the standard deviation of one sample is double that of the other sample, see Table 1 for details). Once the parameters of the 1$^{st}$ sample were generated using the optimization strategy described in the beginning of *Simulations and Results* section, the parameters of the 2$^{nd}$ sample (that produces the same median and skewness but double the standard deviation of sample 1) were computed as:

$$\lambda_{S2} = 2\lambda_{S1}, \ \gamma_{S2} = \gamma_{S1}, \ \delta_{S2} = \delta_{S1}, \ \xi_{S2} = 2\xi_{S1} \tag{12}$$

where S1 and S2 are the 1$^{st}$ and the 2$^{nd}$ samples. The RRO test was performed and the excess type 1 error was computed as in simulation set 1. Figure 2 shows the excess type-1 errors for the RRO test using the Monte-Carlo vs. normal approximation method. For small-sized sample pairs (= 15, 20), the RRO test using the Monte-Carlo method shows lower excess type 1 error as compared to the normal approximation method. The difference also increases as the level of significance decreases. For large equal-sized sample pairs (= 40, 60), the RRO test using the Monte-Carlo method shows higher excess type 1 error. However, the performance of this method converges to that of normal approximation as the level of significance is decreases. As in simulation set 1, the results are similar for equal vs. unequal-sized sample pairs as well as for left vs. right-skewed distributions. An important observation from the results of the unequal-sized sample pairs is that using the normal approximation method, the excess type 1 error of the RRO test is dependent on which sample has a larger dispersion. As can be seen from Figure 2B, when the larger sample has larger dispersion the RRO test using this method performs better than when the smaller sample has larger dispersion. This result is consistent with the findings of Feltovich (see Figure 2 in [6]). However, using the Monte-Carlo method, this variability in performance is not seen.

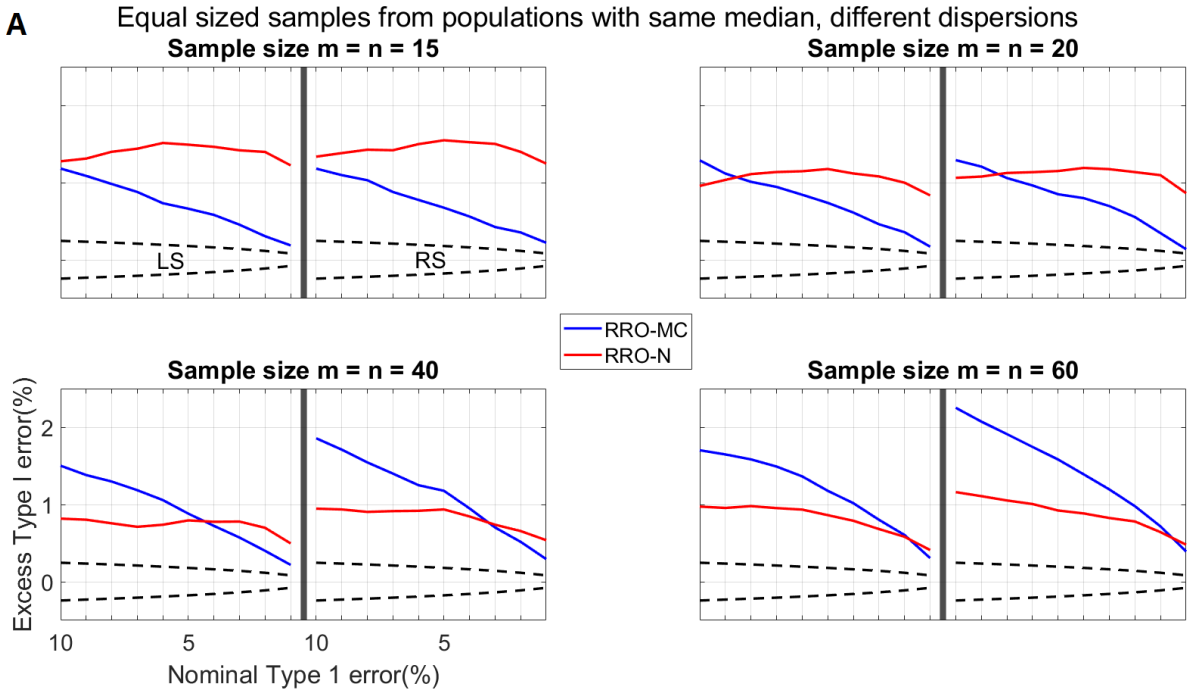

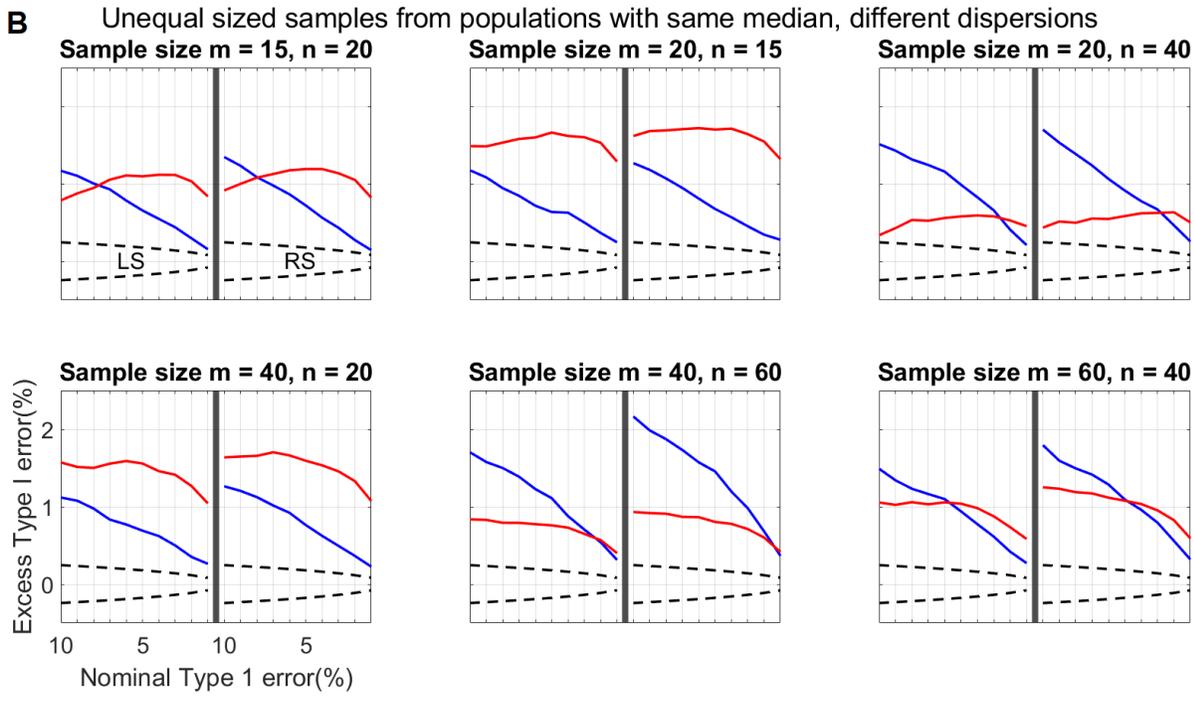

**Figure 2:** Excess type 1 error of the robust rank-order test for sample pairs drawn from the same distribution type (left/right-skewed) with identical median and skewness but different standard deviations (the second sample has a standard deviation twice of that of the first sample, see Table 1 for the values). [A] Equal-sized sample pairs, [B] Unequal-sized sample pairs. The dashed lines enclose the region where the excess type 1 error is not significant (binomial test, p = 0.01). [LS: Left-skewed, RS: Right-skewed, RRO-MC: Robust rank-order test with Monte-Carlo estimation

of test-statistic distribution, RRO-N: Robust rank-order test with standard normal approximation of test-statistic distribution]

**Simulation set 3**

In this simulation set the sample pairs were drawn from the same distribution type (repeated for left-skewed and right-skewed distributions separately, see Table 1) with different medians, but the same standard deviation and skewness (specifically, the second sample had higher median). The difference in central tendencies (i.e. the effect size) was measured using the method proposed by Cliff [12]:

$$d = \frac{\#(S2 > S1) - \#(S2 < S1)}{mn} \quad (13)$$

where $d \in [0, 1]$ is the estimate of effect size, S1 and S2 are the two samples, and # denotes *the number of times*. This measure is robust to unequal sample standard deviations and makes no assumptions about the sample distributions [12]. Using this measure two subsets of simulations were conducted where $d = 0.25$ and $0.5$. In each case, first the parameters of the distributions needed to produce sample 1 with the pre-defined central moments (see Table 1) were generated using the optimization strategy described in the beginning of *Simulations and Results* section. Next, given the optimized parameters of sample 1 and effect size $d$, $\Delta_m$ $(= median[sample\ 2] - median[sample\ 1])$ was computed using MATLAB's global optimization algorithm *patternsearch* (with the inequality constraint $median[sample\ 2] > median[sample\ 1]$). Accordingly, the objective function provided to this algorithm (i) generated a pair of large-sized samples (size = 10,000) with the parameters of sample 1 and, (ii) shifted the 2nd sample by adding $\Delta_m$ to each element. Subsequently the effect size was computed using equation 13 and its L2 error was returned for minimization. After optimization, the sample pairs for the simulation set were generated from the optimized parameters of sample 1 and then the 2nd sample was shifted by adding the optimized value of $\Delta_m$ to each element.

Since we know that the 2nd sample has a higher median as compared to the 1st sample, the RRO test was performed on each sample pair using the Monte-Carlo and normal approximation method and the resultant left-tailed p-values were computed using equation 8b. For each set of 100,000 simulations (of a particular sample-size pair), the power was computed as:

$$power\ (\%) = \left(\frac{n(P < \alpha)}{n(P)}\right) * 100 \quad (14)$$

where $P$ is the set of computed p-values and $\alpha$ is the significance level. Figure 3 shows the power values for the RRO test using the Monte-Carlo vs. normal approximation method. For all cases, the power values were found to decrease with decrease in significance level and there was little difference in the power values obtained by the two methods of RRO.

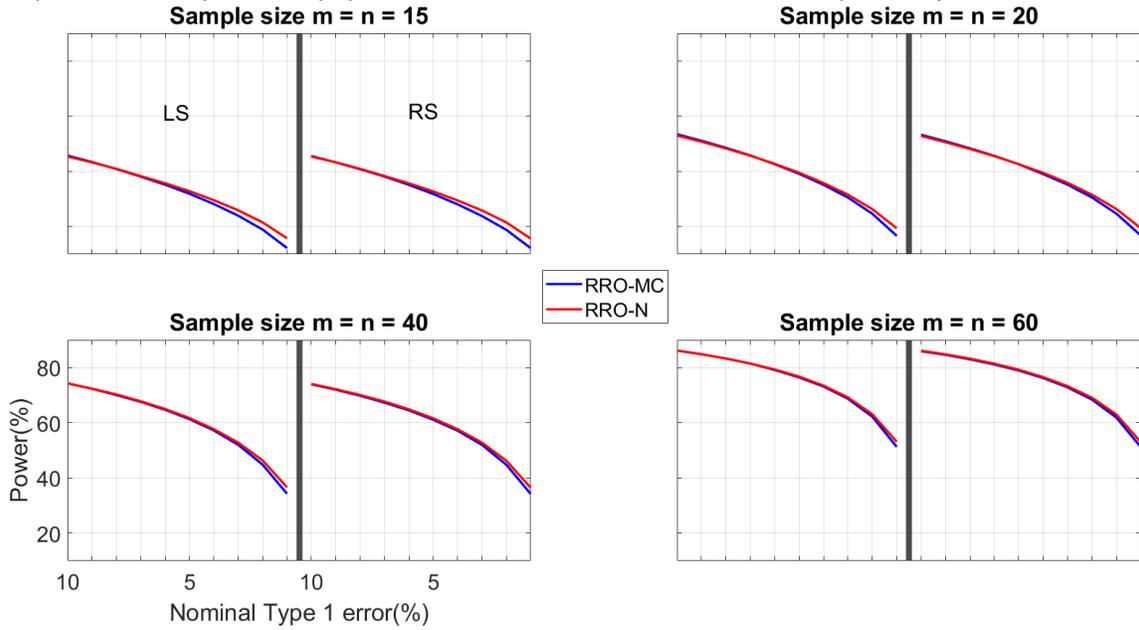
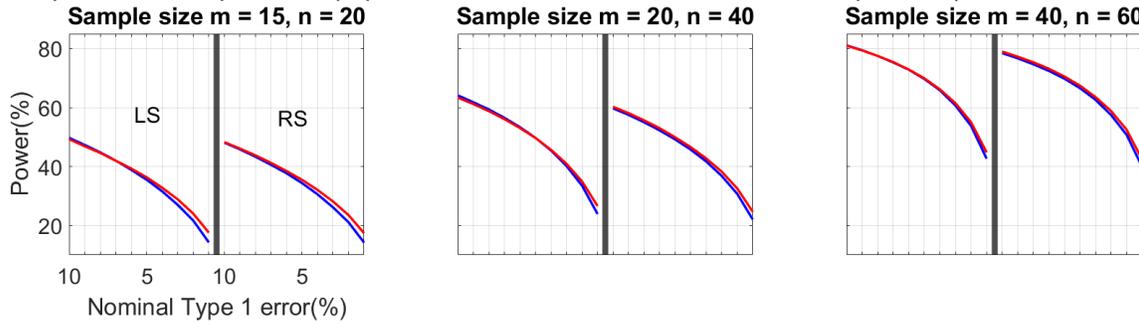

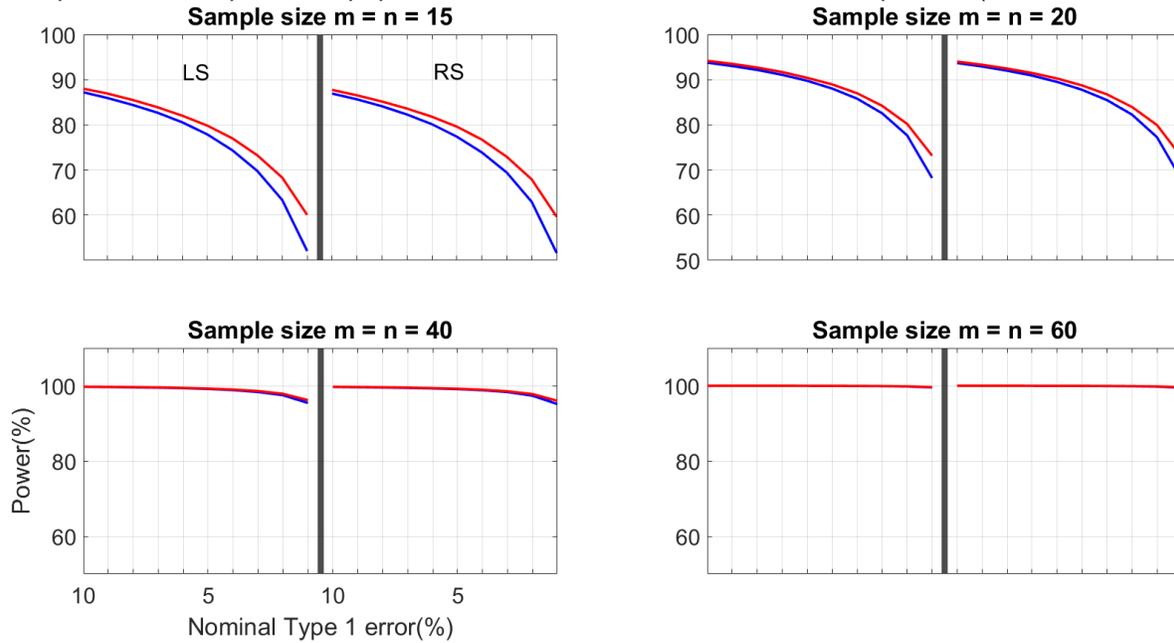

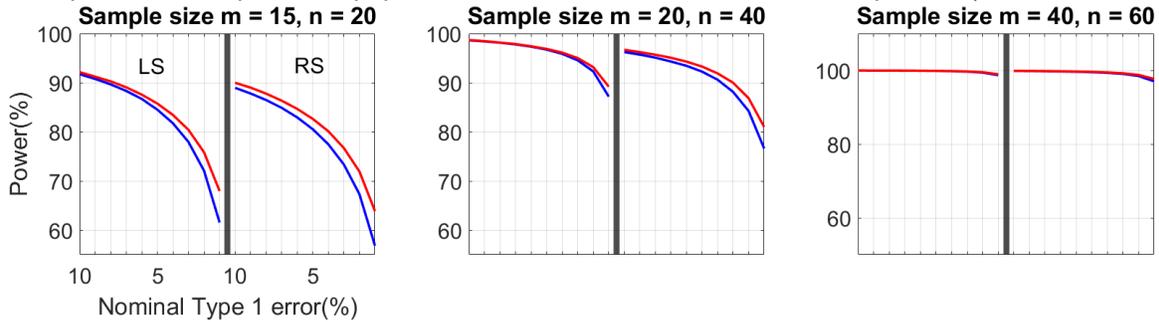

**Figure 3:** Power of the robust rank-order test for sample pairs drawn from the same distribution type (left/right-skewed) with identical standard deviation and skewness but different medians (the second sample has a higher median than that of first sample, see Table 1 for the values). [A] Equal-sized sample pairs, [B] Unequal-sized sample pairs, [C] Equal-sized sample pairs, [D] Unequal-sized sample pairs. The simulations were done for effect size = 0.25 (A and B) and 0.5 (C and D). [LS: Left-skewed, RS: Right-skewed, RRO-MC: Robust rank-order test with Monte-Carlo estimation of test-statistic distribution, RRO-N: Robust rank-order test with standard normal approximation of test-statistic distribution]

**Simulation set 4-7**

For simulation sets 4-7 sample pairs were drawn from distributions whose skewness were varied in addition to their medians and standard deviations. The RRO test was performed for each set and the excess type 1 error was computed as in simulation set 1. For simulation sets 4 and 5, the sample pairs were drawn from distributions with different degrees of right-skewness but with the same median and the same (simulation set 4) or different (simulation set 5) standard deviation(s). Simulation sets 6 and 7 were the same as simulation sets 4 and 5, respectively, except that sample pairs were drawn from distributions with different degrees of left-skewness instead of right-skewness (see Table 1 or values). Figures 4-7 show the excess type-1 errors for the RRO test using the Monte-Carlo vs. normal approximation method for the corresponding simulation sets. For sample pairs with equal standard deviations (simulation set 4 and 6), the observations of simulation set 1 hold true. Likewise, for sample pairs with unequal standard deviations (simulation set 5 and 7), the observations of simulation set 2 hold true. Thus, the performance of the RRO test using the Monte-Carlo method is unaffected by the difference in degree of the skewness of the parent distributions from which the sample pairs are drawn.

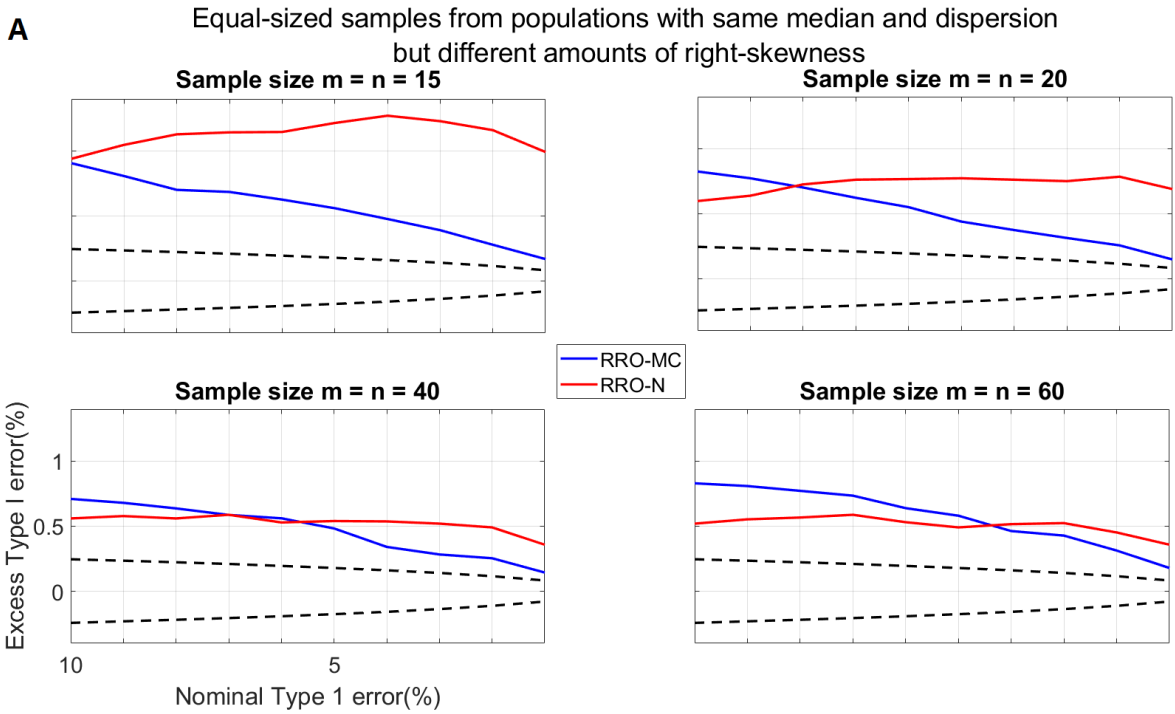

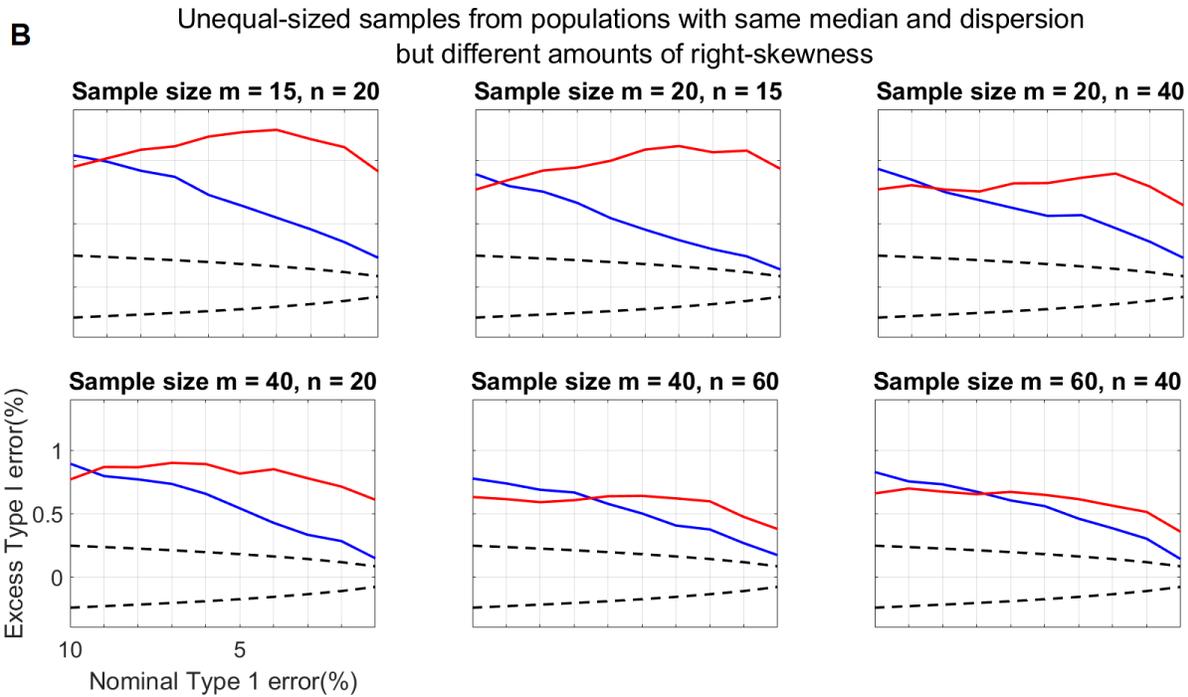

**Figure 4:** Excess type 1 error of the robust rank-order test for sample pairs drawn from distributions with identical median and standard deviation but different degrees of right-skewness (the second sample has higher skewness, see Table 1 for the values). [A] Equal-sized sample pairs, [B] Unequal-sized sample pairs. The dashed lines enclose the region where the excess type 1 error is not significant (binomial test, p = 0.01). [RRO-MC: Robust rank-order test with Monte-Carlo estimation of test-statistic distribution, RRO-N: Robust rank-order test with standard normal approximation of test-statistic distribution]

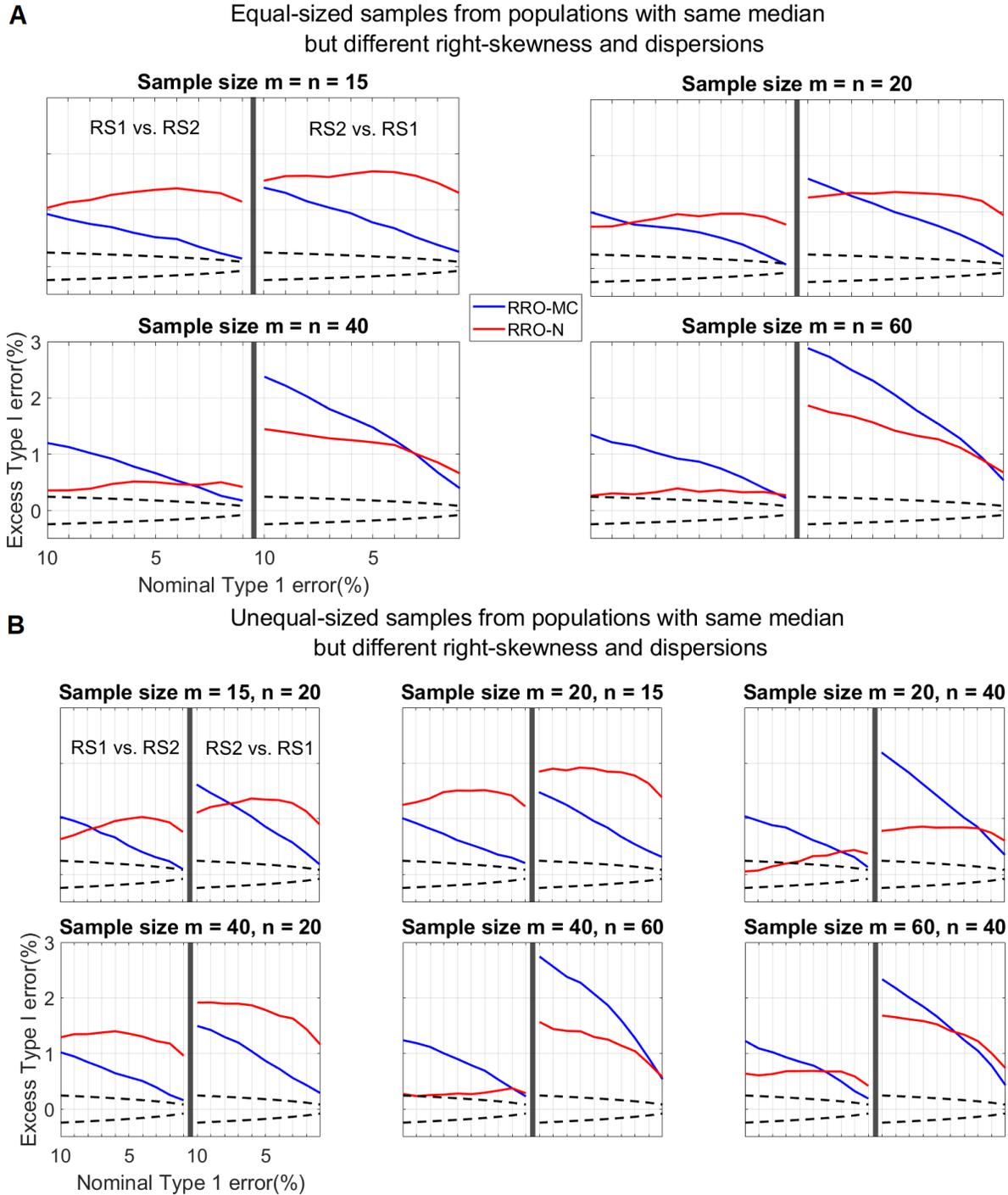

**Figure 5:** Excess type 1 error of the robust rank-order test for sample pairs drawn from distributions with identical median but different right-skewness and standard deviations (the second sample has a standard deviation twice of that of the first sample, see Table 1 for the values). [A] Equal-sized sample pairs, [B] Unequal-sized sample pairs. The dashed lines enclose the region where the excess type 1 error is not significant (binomial test, p = 0.01). [RRO-MC: Robust rank-order test with Monte-Carlo estimation of test-statistic distribution, RRO-N: Robust rank-

test with standard normal approximation of test-statistic distribution, RS1: lower right-skewness (= 1), RS2: higher right-skewness (=1.5)]

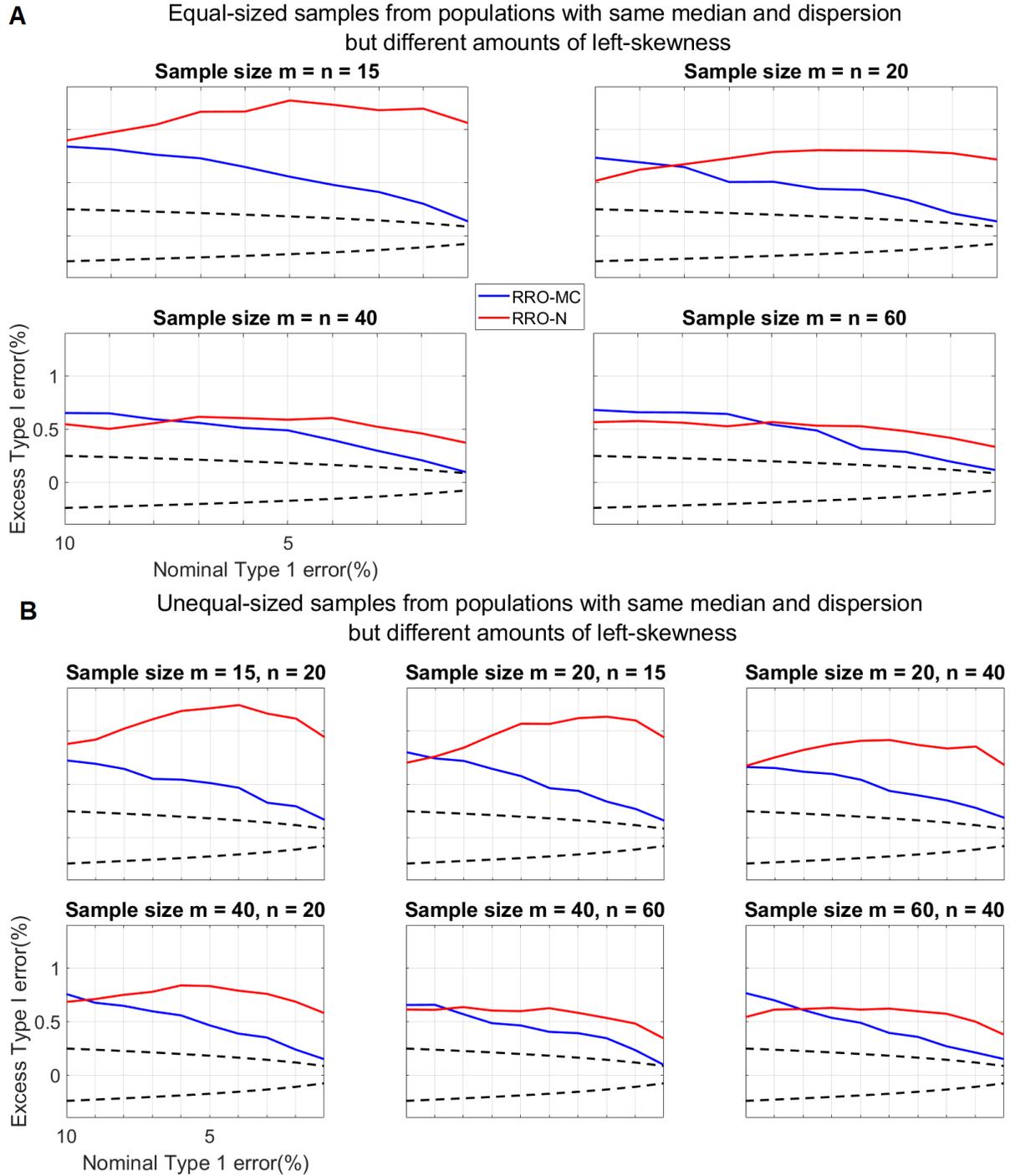

**Figure 6:** Excess type 1 error of the robust rank-order test for sample pairs drawn from distributions with identical median and standard deviation but different degrees of left-skewness (the second sample has higher skewness, see Table 1 for the values). [A] Equal-sized sample pairs, [B] Unequal-sized sample pairs. The dashed lines enclose the region where the excess type 1 error

is not significant (binomial test, p = 0.01). [RRO-MC: Robust rank-order test with Monte-Carlo estimation of test-statistic distribution, RRO-N: Robust rank-order test with standard normal approximation of test-statistic distribution]

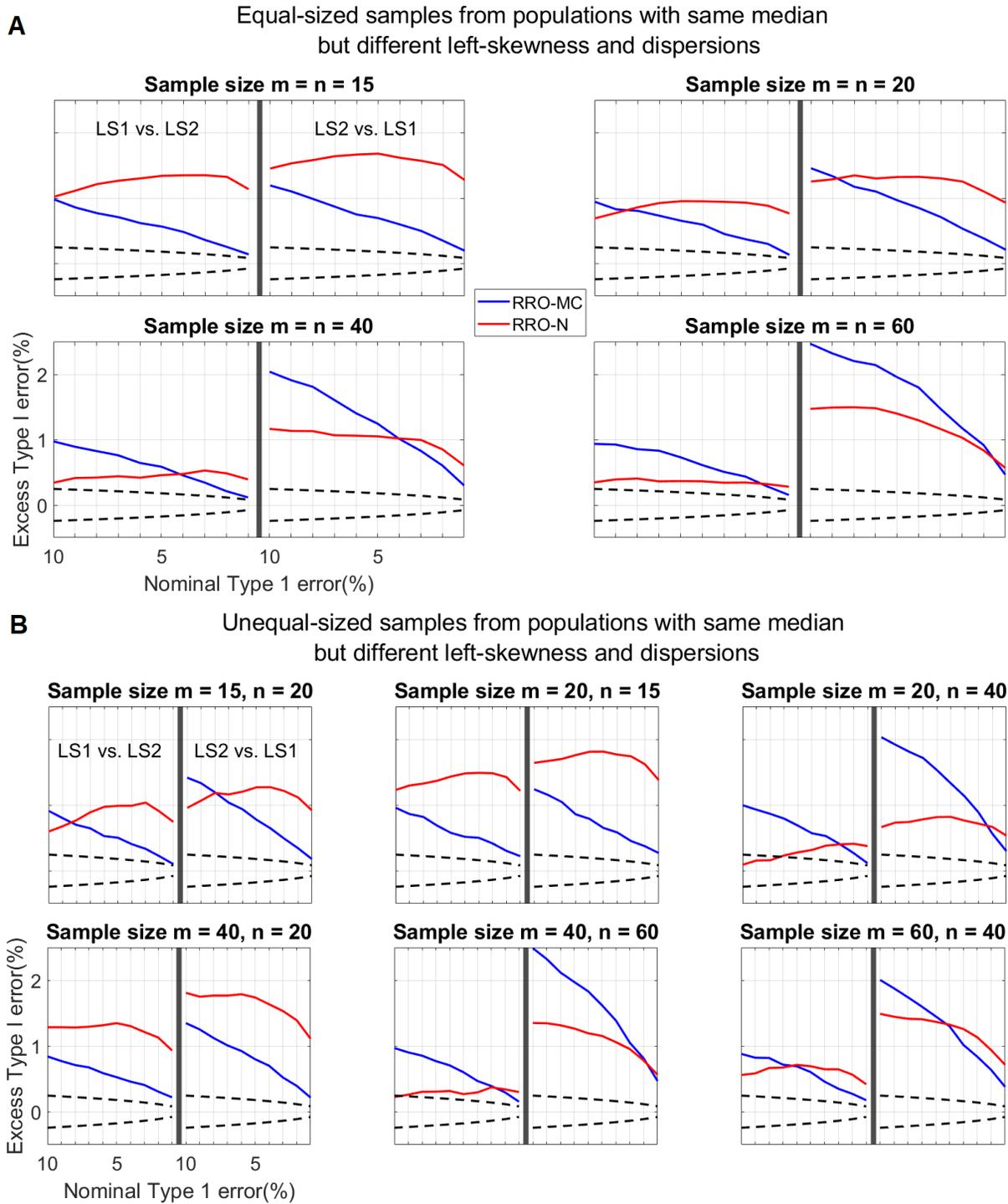

**Figure 7:** Excess type 1 error of the robust rank-order test for sample pairs drawn from distributions with identical median but different left-skewness and standard deviations (the second

sample has a standard deviation twice of that of the first sample, see Table 1 for the values). [A] Equal-sized sample pairs, [B] Unequal-sized sample pairs. The dashed lines enclose the region where the excess type 1 error is not significant (binomial test, p = 0.01). [RRO-MC: Robust rank-order test with Monte-Carlo estimation of test-statistic distribution, RRO-N: Robust rank-order test with standard normal approximation of test-statistic distribution, RS1: lower left-skewness ( = 1), RS2: higher left-skewness (=1.5)]

**Simulation set 8**

Finally, to make the proposed Monte-Carlo method practically useful for calculating p-values on the fly, a simulation was conducted to obtain the minimum number of Monte-Carlo sample pairs for which the proposed method exhibits relatively better performance. Figure 8 shows that the performance of the proposed method asymptotically stabilizes for sample sizes $> 10^4$ for both 5% and 1% significance levels.

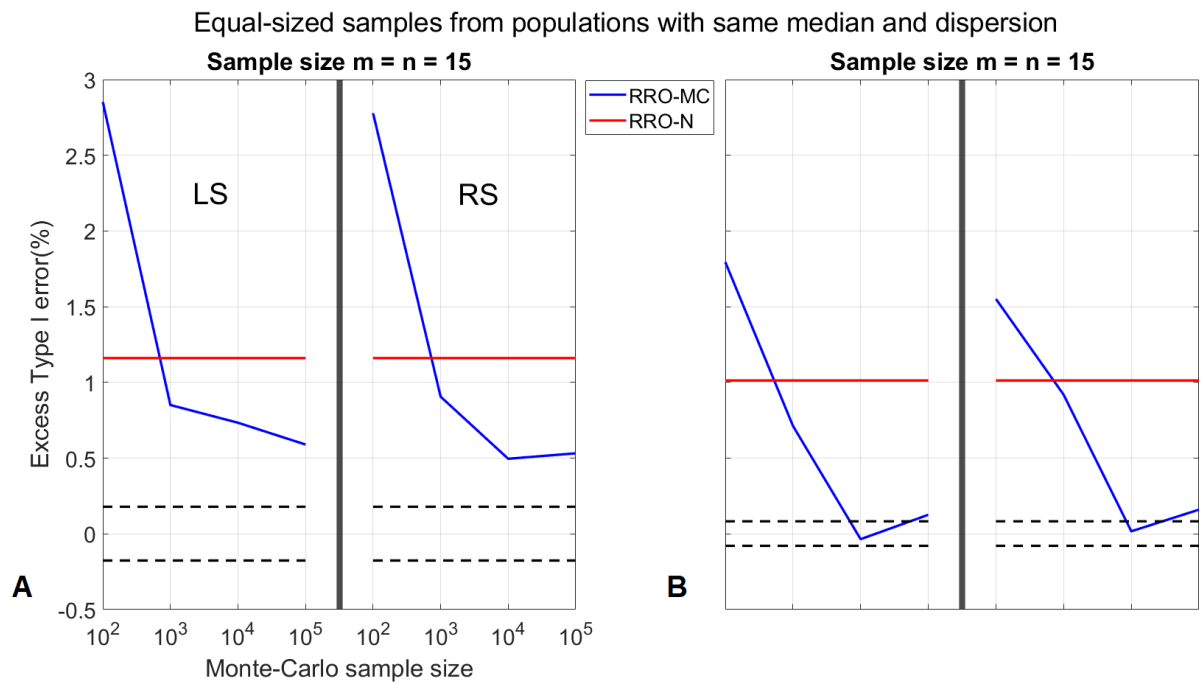

**Figure 8:** Excess type 1 error of the robust rank-order test for sample pairs drawn from the same distribution type (left/right-skewed) with identical median, standard deviation, and skewness for different Monte-Carlo sample sizes. [A] 5% significance level, [B] 1% significance level. The dashed lines enclose the region where the excess type 1 error is not significant (binomial test, p = 0.01). [LS: Left-skewed, RS: Right-skewed, RRO-MC: Robust rank-order test with Monte-Carlo estimation of test-statistic distribution, RRO-N: Robust rank-order test with standard normal approximation of test-statistic distribution]

## Discussion and conclusion

In this study the performance of the RRO test using the Monte-Carlo method to approximate the distribution of the test statistic vs. normal approximation of the distribution of the test statistic was compared using simulations. When the standard deviations of the sample pair were equal, the Monte-Carlo method showed lower excess type 1error, especially for the small-sized sample pairs (= 15,20). When the standard deviations of the sample pair were unequal, for small-sized sample pairs (= 15, 20), the RRO test using the Monte-Carlo method showed lower excess type 1 error as compared to the normal approximation method. The difference also increased as the level of significance decreased. For large equal-sized sample pairs (= 40, 60), the RRO test using the Monte-Carlo method showed higher excess type 1 error. However, the performance of this method converged to that of normal approximation as the level of significance is decreased. Additionally, unlike the normal approximation method, for unequal-sized sample pairs, the performance of the Monte-Carlo method was unaffected by whether the smaller or the larger-sized sample had smaller or larger standard deviation. For effect sizes 0.25 and 0.5 the two methods did not exhibit any difference in power. Moreover, all the above results were consistent for (i) sample pairs drawn from both left and right-skewed distributions, (ii) equal vs unequal-sized sample pairs, (iii) sample pairs drawn from distributions with different degrees of left/right skewness. Finally, a Monte-Carlo sample size of $10^4$ was found to be sufficient to obtain the aforementioned relative improvements in performance.

Based on the results of this study it can be concluded that for small sample sizes ($<= 20$), the Monte-Carlo method outperforms the normal approximation method when comparing samples with asymmetric distributions. This is especially true for low values of significance levels ($< 5\%$). Additionally, when the smaller sample has the larger standard deviation, the Monte-Carlo method outperforms the normal approximation method even for large sample sizes (= 40,60). One caveat is that even though the RRO test using the Monte-Carlo method does show relative improvement over the normal approximation method, the test nevertheless remains liberal i.e. the excess type 1 error is significant for the entire range of significant values examined.

As mentioned in *Methods*, in the real world case, the experimenter does not have prior knowledge of the nature of the parent distribution. Additionally, the method of maximum likelihood fitting is highly dependent on the initial values provided. To address these issues, the author is in the process of developing a MATALB based toolbox which:

a) Obtains the best fit distribution(s) for the given sample pair from a library of standard asymmetric distributions. This step also involves initializing the parameters to be fitted in a way that consistently ensures convergence. For this purpose, it is proposed to start from the analytically calculated/optimized values of the parameters from the 1$^{st}$ 3/4 central moments of the given sample

b) Performs the Monte-Carlo RRO test using these best fit distributions to obtain the p-value estimate.

The need for 2 separate optimizations (at worst) along with the final bootstrapping step raises the question of how practical it will be to use this method for the RRO test on the fly. However, given the powerful processors of modern personal computers and efficient linear algebra techniques in modern programming languages like MATLAB or Python, this author is of the view that a practical use of this method is very much realizable especially when the sizes of the samples being compared are small. In fact, in the real world this is often the case and indeed the proposed method demonstrates its best performance in such cases.